# KAQG: A Knowledge-Graph-Enhanced RAG for Difficulty-Controlled Question Generation

**Ching Han Chen[1], and Ming Fang Shiu[1]**
[1]Department of Computer Science and Information Engineering National Central University, Taiwan

Corresponding author: Ming Fang Shiu (e-mail: 108582003@cc.ncu.edu.tw).

**ABSTRACT** This study introduces Knowledge Augmented Question Generation (KAQG), an educational assessment framework that integrates Item Response Theory (IRT), Bloom's Taxonomy, and knowledge graphs into a multi-agent Retrieval-Augmented Generation (RAG) system. The proposed approach overcomes limitations of existing methods by enabling fine-grained control over item difficulty, psychometric calibration, and cognitive alignment. It employs multi-graph isolation to preserve domain-specific semantics and leverages a distributed agent architecture coordinated through Data Distribution Service (DDS) for scalable and fault-tolerant operations. Each agent specializes in tasks such as retrieval, generation, or evaluation, forming a modular and traceable pipeline. Distinctively, the framework encodes semantic hierarchies, PageRank-based concept weighting, and assessment-theory parameters directly into the generation process, ensuring that questions are both contextually grounded and cognitively calibrated. Deployed at Taiwan's National Institute of Environmental Research, the system has demonstrated practical value by reducing manual workload, improving reliability and validity, and supporting both adaptive and standardized assessments. By integrating psychometric theory with AI-driven retrieval and generation, this work establishes a scalable and cognitively aligned solution for education and professional certification.

## I. INTRODUCTION

The rapid progress of large language models (LLMs) has stimulated their adoption in education, particularly for automated assessment and question generation. However, despite strong language abilities, most current applications remain limited in psychometric rigor, as they rarely address essential requirements such as difficulty control, item calibration, and cognitive alignment for valid educational evaluation.

Retrieval-Augmented Generation (RAG) extends LLMs with domain-specific knowledge and has recently been explored as a foundation for educational applications. GraphRAG [1] improves retrieval accuracy through structured graph representations but does not incorporate mechanisms for cognitive difficulty control. HippoRAG [2] introduces neurobiologically inspired long-term memory for extended reasoning but depends on centralized design, which limits scalability in large-scale assessment contexts. LightRAG [3] emphasizes lightweight efficiency for real-time use, yet it also overlooks psychometric requirements. While these frameworks advance RAG research, they remain insufficient when directly applied to exam question generation, where psychometric validity and cognitive targeting are indispensable.

To address these limitations, this study introduces a novel multi-agent architecture, Knowledge Augmented Question Generation (KAQG), designed specifically for educational assessment. The framework is the first to seamlessly integrate Item Response Theory (IRT) and Bloom's Taxonomy [4] into a RAG pipeline, embedding IRT parameters for precise control of difficulty and discrimination while aligning question generation with hierarchical cognitive levels. By explicitly adapting RAG for assessment purposes, the approach achieves a theoretical breakthrough by bridging psychometric theory with neural generation, while also offering a practical contribution by enabling automated exam systems to produce cognitively calibrated and psychometrically valid questions on a scale.

By applying knowledge graph (KG) techniques, exam question generation can be significantly improved [4]. KGs represent entities and their relations using SPO triples, enabling structured representation and reducing redundancy through entity normalization [6]. Query languages like SPARQL or Cypher support precise retrieval, minimize noise, and enhance inference [7].

To resolve conceptual ambiguity across domains, KAQG employs a multi-graph isolation architecture, where each subject domain is supported by a dedicated knowledge graph [8]. This separation prevents semantic interference, maintains domain-specific clarity, and promotes scalable modularity. These structured KGs provide rich, interpretable semantics that empower RAG with precise context. The integration of KGs enables agentic reasoning and multi-hop retrieval, allowing LLMs to generate coherent, educationally meaningful questions through deterministic logic chains.

To address domain fusion challenges and highlight the synergy of knowledge graphs and RAG, the proposed framework integrates these technologies with assessment theory to enhance question quality and psychometric validity. Unlike approaches that rely solely on graph embeddings, it explicitly encodes semantic types, hierarchical relations, domain ontologies, and formal logic into both retrieval and generation phases [9]. This deep integration ensures that RAG can retrieve contextually relevant content and generate cognitively calibrated questions, aligning structured knowledge with specific educational goals, as illustrated in Figure 1.

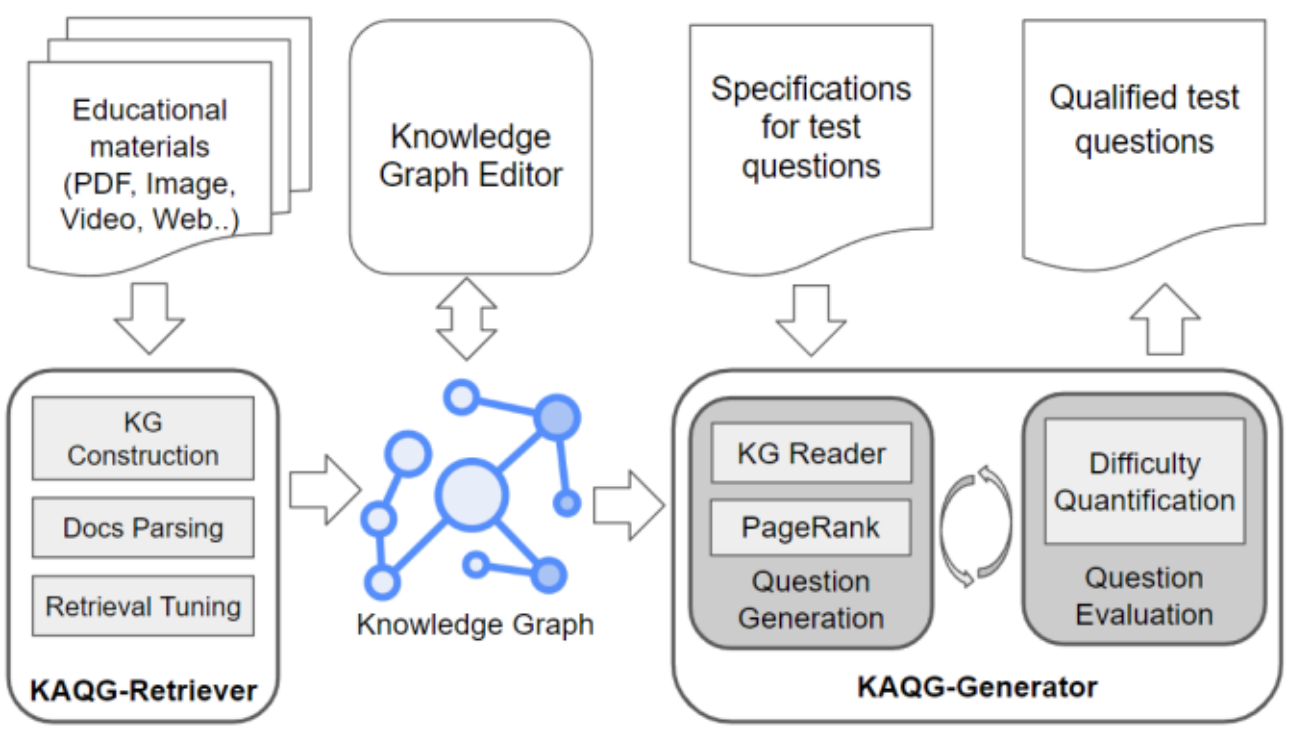

**FIGURE 1. The KAQG system architecture for question generation.**

### A. KAQG-RETRIEVER

This module offers a scalable method for converting educational content into structured, domain-specific knowledge graphs using LLMs [10]. By isolating subjects, it ensures semantic clarity and avoids cross-domain interference. Agentic reasoning enhances retrieval through dynamic query refinement and task delegation, enabling the extraction of validated knowledge for standardized test generation.

### B. KAQG-GENERATOR

This component transforms structured knowledge into psychometrically valid questions by integrating knowledge graphs with assessment theory. Using PageRank to rank key concepts [11], LLMs generate items aligned with Bloom's Taxonomy and validated via IRT. Agentic reasoning supports autonomous retrieval and coordination, ensuring coherent, curriculum-based questions for large-scale testing [12].

### C. AI AGENTS FRAMEWORK

KAQG adopts a distributed AI agent architecture for scalable, autonomous exam generation [13]. With DDS coordination, agents manage tasks in parallel to avoid bottlenecks. Specialized LLM agents adapt content to various standards, offering a practical solution for institutions aiming to automate validated test creation at scale [14].

The system was deployed at Taiwan's National Environmental Research Academy, where it produced standardized exam items, reduced manual effort, and improved domain adaptability through agent-based collaboration and structured reasoning. Code is available at https://github.com/mfshiu/kaqg, with the agent framework at https://github.com/mfshiu/AgentFlow.

## II. METHODS

This chapter introduces an integrated framework that uses domain knowledge to generate high-quality exam questions. It consists of two components: a retriever that builds knowledge graphs and processes educational content, and a generator that ranks concepts via PageRank, formulates questions, and evaluates their difficulty and quality.

### A. KAQG-RETRIEVER: KNOWLEDGE GRAPH EXTRACTION

The retriever transforms multimodal educational materials—such as audio, video, images, PDFs, and text—into structured knowledge graphs. It first converts all inputs into textual form suitable for NLP processing [15], then employs LLMs to extract key entities, relationships, and hypernyms as structured triples. These triples are merged with concept mappings and hierarchical structures to form a consolidated graph. Each subject is stored in a dedicated knowledge graph to ensure semantic relevance and prevent cross-domain interference, laying a solid foundation for applications such as question generation and intelligent tutoring [16].

Let $D = \{d_1, d_2, \cdots, d_n\}$ represent the set of all source documents, where each $d_i$ can be of various formats (audio, video, PDF, or text). We define a transcription function

$$\mathcal{T}: \mathcal{D} \rightarrow \mathcal{S}. \qquad (1)$$

Such that for every document $d_i \in \mathcal{D}$, we obtain a corresponding text $s_i = \mathcal{T}(d_i)$. This process may involve automatic speech recognition (ASR) for audio/video inputs or optical character recognition (OCR) for scanned PDFs. As a result, we accumulate a set of textual documents $S = \{s_1, s_2, \cdots, s_n\}$ that can be further processed with NLP methods.

Once we have the textual representations, we employ a large language model or other NLP techniques to convert each text $s_i$ into a set of extracted triples. Formally, we define an extraction function

$$\mathcal{E}: \mathcal{S} \rightarrow \mathcal{P}(\mathbb{T}). \qquad (2)$$

Where $\mathbb{T}$ is the set of all possible triples, and $\mathcal{E}(s_i)$ returns a subset of these triples. Here, $\mathcal{P}(\mathbb{T})$ denotes the power set of $\mathbb{T}$, which means the set of all possible subsets of triples that could be extracted from text. Each triple $(h, r, t)$ consists of a head entity $h$, a relation $r$, and a tail entity $t$. This extraction step includes: (1) entity recognition to identify textual entities, (2) relationship detection to capture any relevant associations, and (3) attribute extraction for entity properties. Moreover, for each recognized entity, we map it to a higher-level concept entity (i.e., a hypernym) in order to provide hierarchical context.

Our framework defines three core types of nodes in the knowledge graph:

1) textual nodes $N_{\text{text}}$, extracted directly from content.
2) concept nodes $N_{\text{concept}}$, representing key domain entities.
3) hierarchical nodes $N_{\text{hierarchy}}$, aligned with directory.

To link textual nodes to concepts, we define a concept mapping function:

$$\mathcal{C}: N_{text} \rightarrow \mathcal{P}(N_{concept}). \quad (3)$$

This maps each textual node to one or more corresponding concept nodes, enabling semantic normalization and reducing redundancy. For example, both "mercury" and "Hg" are mapped to the unified concept of "chemical element."

We further organize concepts using hierarchical structures $N_{\text{hierarchy}} = \{h_1, h_2, \cdots\}$, and define two types of directed edges to support structured reasoning and modular graph expansion:

1) $(h_j, \rho_{\text{part-of}}, h_i)$: denotes hierarchical nesting among directory nodes.
2) $(c,\ \rho_{\text{include-in}}, h)$ : links a concept node $c \in N_{\text{concept}}$ to a directory node $h \in N_{hierarchy}$.

To associate a textual node $e \in N_{\text{text}}$ with its semantic concept $c \in N_{\text{concept}}$, we define an "is a" relation $\rho_{\text{is-a}}$, forming triples of the form $(e,\ \rho_{\text{is-a}}, c)$.

Finally, the knowledge graph $G$ is formulated as a directed graph $G = (N, E)$, where the node set is:

$$N = N_{text} \cup N_{concept} \cup N_{hierarchy}. \quad (4)$$

And the edge set is:

$$E = E_{triplet} \cup E_{concept-map} \cup E_{hierarchy}. \quad (5)$$

Here,

1) $E_{triplet}$ contains semantic relation triples (e.g., (h, r, t)) extracted from text.
2) $E_{concept-map}$ includes "is-a" links between textual and concept nodes.
3) $E_{\text{hierarchy}}$ represents directory nesting and concept-to-directory associations.

This structured graph design enables unified reasoning across raw text, abstract knowledge, and curricular hierarchies.

A sample result of the knowledge graph is shown in Figure 2. This unified framework not only supports efficient querying and reasoning over diverse data but also enables advanced analytics, such as semantic search and entity disambiguation, paving the way for more intelligent knowledge-driven applications [17].

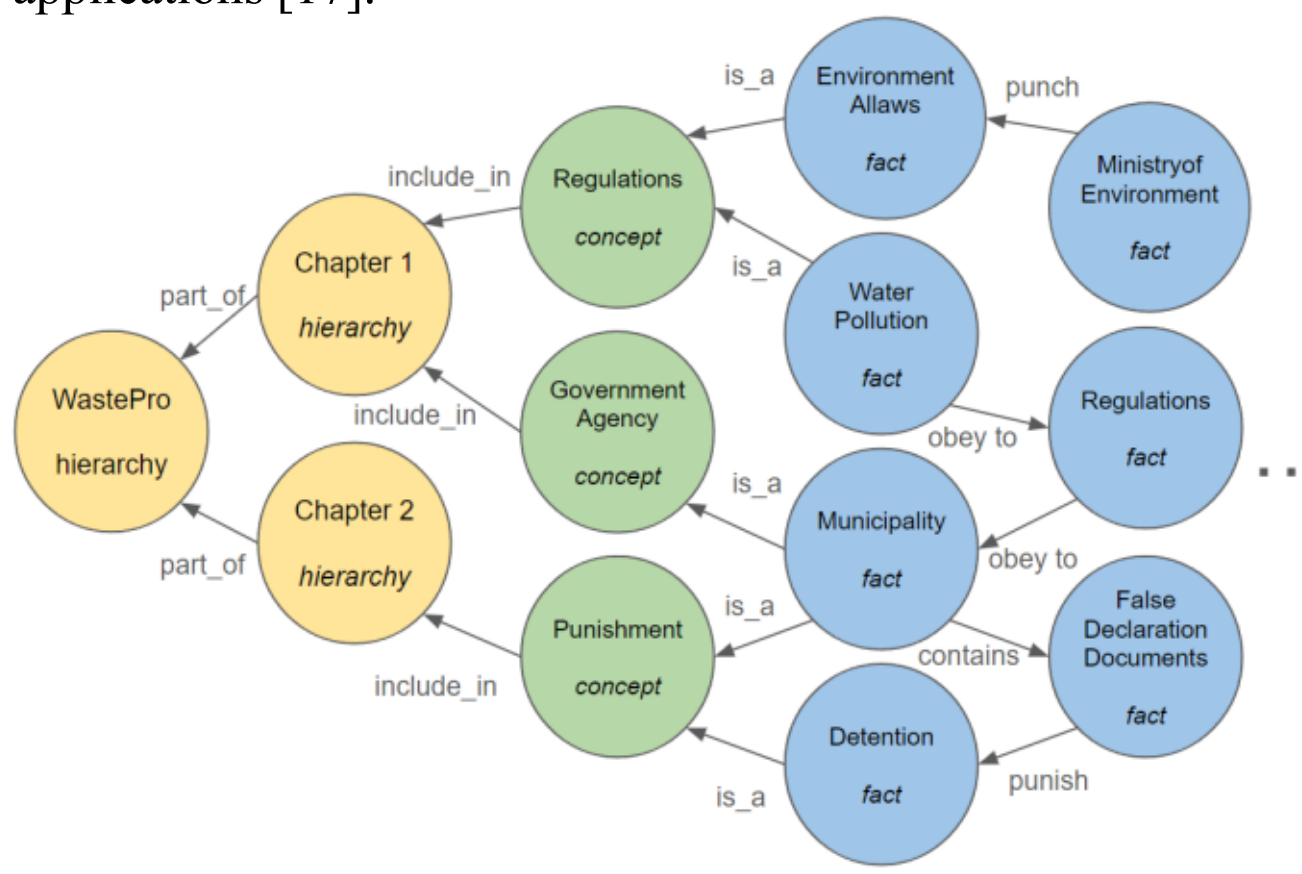

**FIGURE 2.** **A scenario of a knowledge graph derived from a parsed document.**

To mitigate potential errors from LLM-based extraction, we implemented entity normalization using concept mappings and hierarchical ontologies. However, we recognize the need for formal benchmarking. Ongoing efforts are being made to construct labeled datasets and apply inter-annotator agreement for validation, especially when extracting from noisy multimodal inputs like videos and scanned documents.

### *B. KAQG-RETRIEVER: MULTIMODAL INTEGRATION AND CONCEPT MAPPING*

This module applies a two-stage process—multimodal unification and concept harmonization—to integrate heterogeneous sources into a coherent knowledge graph, ensuring semantic consistency and reliability in subsequent tasks, as illustrated in Figure 3.

**Knowledge Graph Construction and Validation**

1) **Graph establishment and validation:** Knowledge graphs are built by applying an LLM-based entity–relation extraction pipeline. Each triple is stored with provenance tags (e.g., textbook, transcript, web page) to ensure source traceability. Validation is performed through three complementary mechanisms: (a) entity normalization against domain ontologies, (b) rule-based conflict resolution, and (c) expert inspection of sampled triples for accuracy.
2) **Concept mapping:** Textual nodes are first analyzed by an LLM, which identifies candidate concepts, hypernyms, and semantic roles from context. These LLM-derived concepts are then normalized through a mapping function that combines (a) rule-based synonym and abbreviation resolution with (b) ontology-driven alignment. For example, the symbol "Hg" detected in a scanned PDF and the term "mercury" extracted from a video transcript are jointly mapped into the unified concept node *chemical element*. This hybrid approach ensures that concepts are simultaneously inferred from LLM contextual analysis and validated against structured ontologies, balancing flexibility with semantic rigor.

**Multimodal Data Integration**

1) **Preprocessing pipeline:** Inputs from various sources are systematically transformed into textual representations:
   1. PDFs / scanned documents → Optical Character Recognition (OCR)
   2. Audio / video lectures → Automatic Speech Recognition (ASR)
   3. Images or diagrams → Captioning or text extraction using vision-language APIs
   4. Web content → HTML parsing and boilerplate removal

   The resulting unified corpus is processed through the LLM-based extraction module to generate standardized triples with node types (textual, concept, hierarchical). Provenance tags are preserved across all formats.

2) **Evaluation of conversion quality:** Conversion quality is assessed using standard tools and APIs (e.g., Tesseract OCR, Whisper ASR). Performance is verified via spot-checking recognition accuracy, inter-annotator agreement on extracted triples, and error analysis of noisy inputs (e.g., low-resolution scans or domain-specific jargon). Identified challenges include handling ambiguous terminology and imperfect transcription, which are mitigated by expanding the ontology and refining LLM prompts for entity disambiguation.

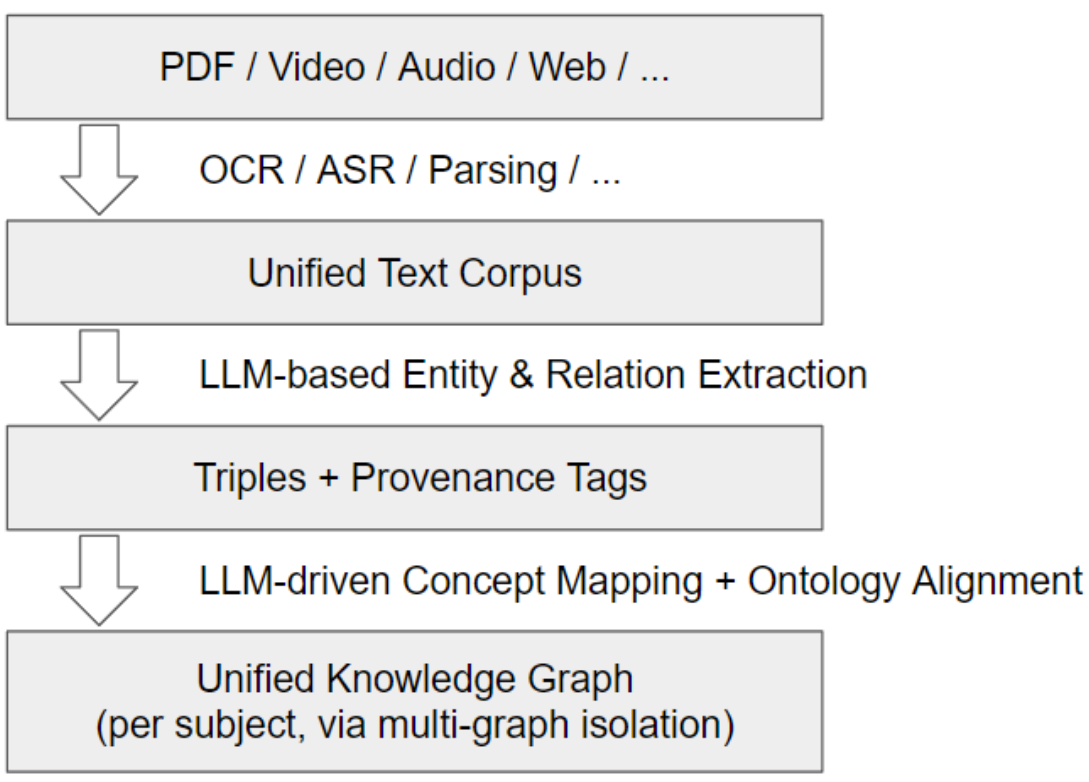

**FIGURE 3.** Workflow of KAQG-Retriever for multimodal integration and concept mapping.

The workflow normalizes multimodal sources into text and harmonizes them into unified concept structures. Through LLM-based analysis, synonym handling, and ontology alignment, heterogeneous materials are scalably integrated to support accurate retrieval and cognitively calibrated question generation.

### C. KAQG-RETRIEVER: CROSS-DOMAIN ISOLATION

To avoid knowledge interference across exam subjects, our approach adopts a multi-graph architecture, where each subject $a_i \in A = \{a_1, a_2, \cdots, a_n\}$ is assigned its own standalone knowledge graph $G_i = (V_i, E_i)$. This design preserves domain purity, ensuring that content and terminology from one subject do not contaminate or distort another. It also facilitates modular updates and enables independent graph evolution.

Unlike a monolithic graph that grows complex and unmanageable as domains expand, this multi-graph strategy simplifies maintenance, supports discipline-specific refinement, and allows subject matter experts to govern their graphs without cross-domain interference [18]. As a result, retrieval becomes more accurate, scalable, and adaptable to domain-specific tools and methods.

### D. KAQG-GENERATOR: THEORY-DRIVEN QUESTION GENERATION

The generator automates the creation of exam questions through a structured workflow integrating knowledge graphs and language models. By leveraging concept hierarchies, semantic relationships, and structural signals, the system aligns generated questions with varying cognitive demand while maintaining consistency and relevance across content.

LLMs are used for ranking, transforming, and phrasing candidate items, while validation techniques ensure accuracy and coherence [19]. This theory-driven design not only improves logical structure and pedagogical alignment but also supports flexible adaptation to diverse assessment needs.

In the three-parameter logistic (3PL) model of IRT, the probability of a correct response is defined as:

$$P_{correct} = c + \frac{1 - c}{1 + e^{-a(\theta - b)}} \tag{6}$$

Here, $\theta$ denotes the test taker's latent ability, while $a$, $b$, and $c$ represent discrimination, difficulty, and guessing parameters. Together, they model how items differentiate proficiency, control challenge level, and filter random success.

In system, item parameters are calibrated using Bloom's Taxonomy and knowledge graph signals:

1) **Difficulty ($b$)**
   Higher-order tasks (e.g., Analyze, Evaluate) align with concepts of higher PageRank or deeper hierarchy, indicating more abstract or central content and thus higher difficulty.
2) **Discrimination ($a$)**
   Items based on well-connected or deep concepts better distinguish learner ability by targeting nuanced understanding.
3) **Guessing ($c$)**
   Multi-hop reasoning paths reduce random guessing, lowering the c parameter and enhancing assessment integrity.

This graph-driven approach ensures that questions are both semantically grounded and cognitively aligned [20].

### E. KAQG-GENERATOR: DEFINING DIFFICULTY LEVELS

The generator controls question difficulty by evaluating seven surface features of multiple-choice questions: stem length, domain-specific vocabulary, cognitive demand, option length, option similarity, stem-option overlap, and distractor plausibility. Each feature is rated on a 3-point scale—low (1), medium (2), high (3)—based on its contribution to cognitive complexity and assessment quality. This structure enables automatic difficulty classification, ensuring pedagogical alignment and empirical calibration.

Once each feature is assigned a difficulty rating (1, 2, or 3), the total difficulty score Q of the MCQ is the sum of these feature scores, as shown in the following model:

$$Q = \sum_{i=1}^{7} f_i \tag{7}$$

where $d_i \in \{1, 2, 3\}$, is the difficulty score for the *i*-th feature. This straightforward summation integrates multiple cognitive and linguistic elements into a single metric, supporting consistent item difficulty classification across different assessments.

Building on this straightforward summation, not all features exert an equal impact on item difficulty, so weighing each attribute addresses variations in their influence [21]. For example, advanced cognitive demand generally affects difficulty more than simpler textual factors. By assigning a weight $w_i$ to each feature i, drawn from expert judgment or statistical calibration, we adjust the total difficulty to:

$$Q_{weighted} = \sum_{i=1}^{7} w_i \cdot f_i \tag{8}$$

where $f_i \in \{1, 2, 3\}$, is the difficulty level for each feature. The weights $w_i$ were derived through a structured expert-rating process. Experts rated the relative impact of each feature (e.g., cognitive level, distractor plausibility) on item difficulty and discrimination. Final weights were computed by normalizing these ratings. While not yet statistically calibrated, this method provides a solid theoretical foundation, with future work aimed at refining these weights through statistical calibration.

Beyond expert-assigned weights, the framework incorporates IRT parameters (cf. (6)) to anchor feature contributions in psychometric evidence. Feature weights are dynamically adjusted as

$$w_i^* = w_i \cdot (1 + \alpha \cdot a) + \beta \cdot b, \tag{9}$$

where *a* and *b* are the discrimination and difficulty parameters, respectively, and α, β are scaling constants. Higher Bloom levels (e.g., Analyze, Evaluate) are associated with larger *a*, thus amplifying their influence, while *b* provides statistical grounding for difficulty calibration. The final adaptive score is

$$Q^* = \sum_{i=1}^{7} w_i^* \cdot f_i \tag{10}$$

ensuring that linguistic and structural features are simultaneously normalized by expert judgment and anchored in psychometric validity.

### F. KAQG-GENERATOR: QUESTION GENERATION VIA LLMS

We begin from the perspective of an institution tasked with designing a comprehensive test for a specific subject, leveraging relevant textbooks as foundational content. In academic settings, test creation must ensure coverage of the curriculum's breadth while maintaining alignment with standardized guidelines. To achieve this, the first step often involves defining a distribution ratio for each chapter or topic within the textbook. Mathematically, if $\alpha_i$ denotes the fraction of total questions devoted to chapter $i$, and $n_i$ represents the desired number of questions from chapter iii, then the ratio can be formulated as:

$$\alpha_i = \frac{n_i}{\sum_{j=1}^{k} n_j} \tag{11}$$

where $k$ is the total number of chapters or major divisions of the subject matter. This approach ensures that each section is proportionally represented according to institutional priorities and learning objectives.

Once chapter-specific ratios are established, the test-design process proceeds with identifying crucial concepts from each chapter—typically through a structured knowledge base or knowledge graph. Importantly, the concept nodes extracted in the previous stage can now play a key role in this generation phase, as they serve as foundational anchors for further knowledge expansion. Each key concept can then be expanded by exploring related nodes, such as prerequisite relationships or hierarchical links [22]. Crucially, concepts are ranked based on their instructional value using algorithms like PageRank. For any knowledge point *v*, the PageRank score PR(*v*) is updated iteratively by

$$\mathrm{PR}(v) = (1 - d) + d \sum_{u \in \mathrm{In}(v)} \frac{\mathrm{PR}(u)}{|\mathrm{Out}(u)|} \tag{12}$$

where In(*v*) represents inbound links to *v*, Out(*u*) denotes outbound links from *u*, and $0 < d < 1$ is a damping factor. Higher PageRank scores guide the selection of more essential topics, thus ensuring that the resulting questions cover the subject's core content.

All fact nodes connected to a concept via an *is_a* relationship is also evaluated using the same PageRank algorithm to determine their importance. For each concept, all significant facts—along with their sub-connections—are subsequently incorporated as material for question generation, thus enriching the depth and breadth of the test items.

To ensure cognitive alignment, the system defines prompt templates and control tags corresponding to levels such as "Understand," "Apply," "Analyze," and "Evaluate." For example, questions involving cause-effect reasoning are tagged as "Analyze," while those requiring judgment of arguments are marked "Evaluate." These tags guide the LLM to generate questions with appropriate cognitive depth.

Each concept is assigned a level based on its instructional role and graph centrality. The system then selects a matching template, ensuring that generated questions reflect the intended mental operations rather than surface features—enhancing both pedagogical clarity and assessment validity.

### G. KAQG-GENERATOR: EVALUATION OF GENERATED QUESTIONS

Within this framework, the Evaluation Agent ensures each generated question aligns with the target difficulty level. By leveraging established testing principles, the system assesses item features—such as cognitive load, prerequisite knowledge, and reasoning complexity—and computes a difficulty score through a weighted combination of these features. Let $Q$ represent the resulting difficulty score. To confirm alignment with an institution's intended difficulty $Q^*$, the *Evaluation Agent* verifies whether the absolute difference $|Q - Q^*|$ falls within a predefined tolerance ε. A smaller deviation implies that the item is well-calibrated, while a larger deviation may necessitate revising feature weights or modifying question content. In general, a higher weight yields a more difficult item, whereas a lower one results in an easier question.

### H. AI AGENTS FRAMEWORK: DISTRIBUTED AND AUTONOMOUS OPERATIONS

Generating exam questions from educational materials requires integrating domain knowledge with retrieval, reasoning, and evaluation. Traditional monolithic systems face bottlenecks and lack fault tolerance due to centralized design, making them ill-suited for complex educational tasks. KAQG addresses this via a Multi-Agent System (MAS) that distributes tasks among autonomous agents—for retrieval, generation, and evaluation—enabling parallelism, scalability, and resilience [23].

To coordinate agent interactions, the system employs a Data Distribution Service (DDS) using a publish-subscribe model [24]. DDS decouples agents, allowing real-time coordination, dynamic discovery, and fault-tolerant behavior. This enables agents to operate independently and join or leave without disrupting the system, thus enhancing modularity and robustness, as illustrated in Figure 4.

In this workflow, once a prerequisite task is completed, the responsible agent publishes to a specific topic, and downstream agents subscribed to that topic automatically begin processing, thereby completing the retrieval → generation → evaluation pipeline. To improve throughput, the system also allows multiple agents with identical functions to bid for a task, implementing cluster-based load balancing that prevents redundant outputs and enhances efficiency.

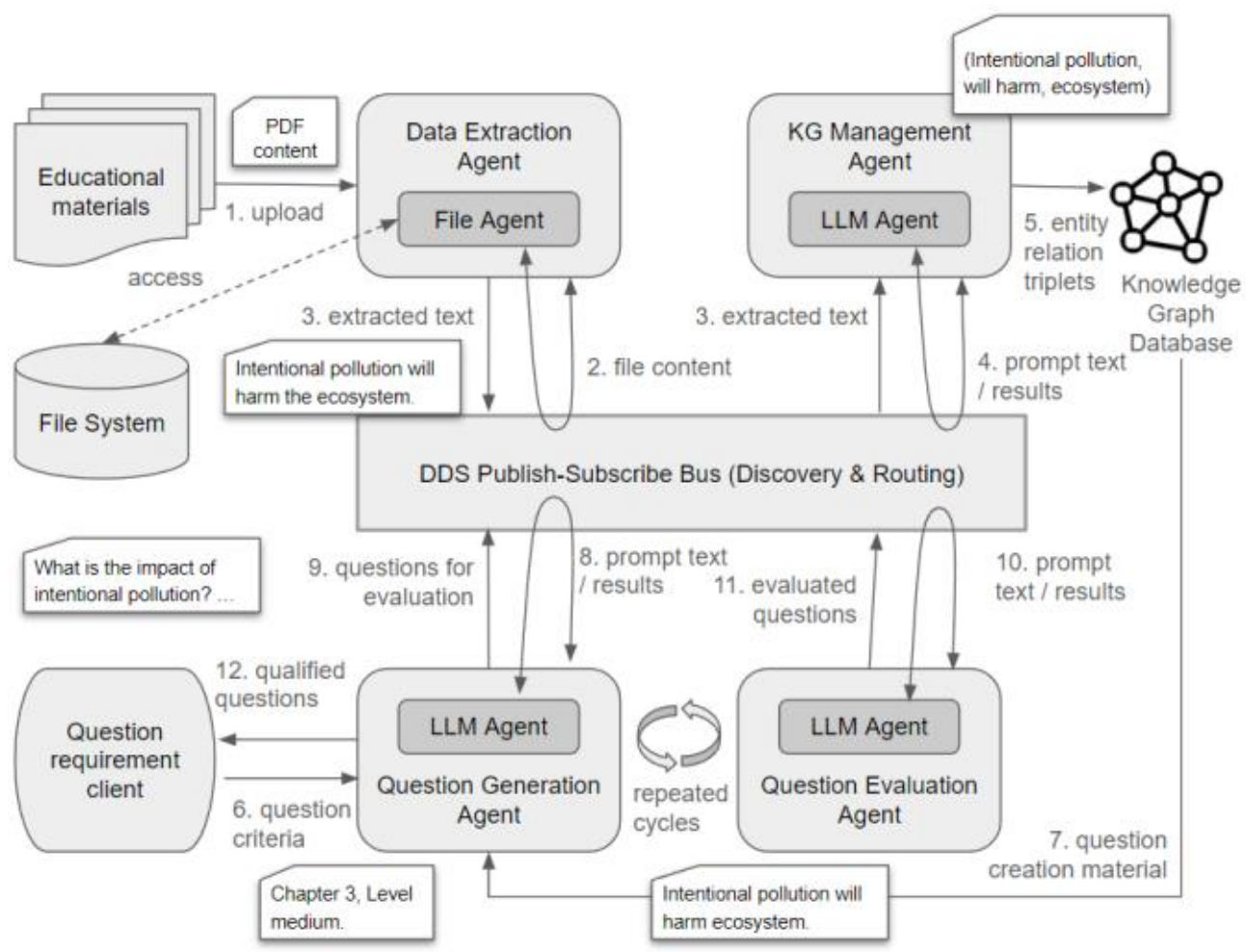

**FIGURE 4. The diagram represents the multi-agent architecture of KAQG with DDS-based publish-subscribe communication.**

The process starts with uploading educational content, parsed by Data Extraction and File Agents. Extracted entity-relation triples are stored in the Knowledge Graph by the KG Management Agent. The Question Generation Agent, supported by the LLM Agent, creates questions using both semantic structure and contextual signals. Agent-based reasoning improves multi-hop inference and domain understanding, while distributed operations increase scalability. The Question Evaluation Agent reviews outputs, which are then filtered and sent to the Question Requirement Client.

## III. EXPERIMENTS

In addition to evaluating item quality, we employ two complementary protocols to examine whether the proposed graph-based reasoning and difficulty modeling achieve psychometric alignment: an ACT-based human benchmarking study (Sec. A–E) and a fully computer-run simulation (Sec. F) that quantifies the specific contributions of IRT weighting and Bloom-level alignment through parameter recovery and ablation. Together, these protocols isolate the effects of assessment-theory integration on difficulty control, item calibration, and cognitive alignment, providing converging evidence of the framework's measurement validity.

### A. EXPERIMENT OBJECTIVES

The ACT (American College Test) is a widely used standardized exam for U.S. college admissions. Its Reading section includes passages with multiple-choice questions to assess comprehension, inference, and analysis. Due to its reliability, the ACT serves as a strong benchmark for evaluating automatically generated reading items.

This experiment tests whether the system—configured at high, medium, and low difficulty—can produce items comparable to official ACT questions. It also validates the impact of novel components, such as difficulty control and multi-step reasoning, by comparing generated items under various conditions.

The study examines three metrics: (1) item difficulty (P value), (2) discrimination index (item effectiveness in differentiating performance), and (3) expert-rated item quality. Improvements across these metrics would confirm the necessity and value of the system's innovative modules, including knowledge graphs and assessment theory integration.

Three ACT passages (A, B, C) were selected, each with 10 official items (30 total as control). Using the same passages, the system generated 90 experimental items (10 per passage per difficulty level). This setup allows direct comparison to demonstrate the contribution and necessity of each innovation. All items and answers are listed in Appendix A.

### B. EXPERIMENTAL PROCEDURE

In this study, we recruited a sufficient number of participants (ideally several dozen, though the exact count varied according to available resources) to ensure reliable statistical outcomes. Participants were then divided into four groups, with each group answering a different set of questions:

1) **Group ACT**
   This group completed the 30 official ACT items, consisting of 10 questions for each passage (A, B, and C).
2) **Group Low**
   This group answered the 30 system-generated items set at low difficulty.
3) **Group Medium**
   This group answered the 30 system-generated items set at medium difficulty.
4) **Group High**
   This group answered the 30 system-generated items set at high difficulty.

Participants read the same three passages (A, B, C) but answered only the questions assigned to their difficulty group or the official ACT set. All items were multiple-choice with one correct answer, delivered via paper or an online platform (the latter allowing response time logging). After testing, responses were submitted for analysis. Item difficulty (P value) was calculated as the proportion of correct answers, while the discrimination index was computed by comparing performance between high- and low-scoring subgroups.

### C. EVALUATION METRICS AND STATISTICAL ANALYSIS

To assess question quality, we compared system-generated items to official ACT questions using statistical analysis across four groups.

1) **Difficulty (P Value)**
   The P value, the proportion of correct responses, was averaged for each group (ACT, Low, Medium, High) across passages A, B, and C. Lower values indicate harder items. This analysis tested whether system-defined difficulty levels yield distinct ranges and how they compare to ACT benchmarks.
2) **Discrimination Index**
   We measured how well each question distinguished high- and low-performing participants by calculating the difference in correct response rates between the top and bottom quartiles. Higher values indicate better discrimination. This was used to compare system-generated items (Low, Medium, High) with ACT questions.
3) **Expert Ratings**
   Experts rated each question's clarity, relevance, and distractor quality on a 5-point scale. These ratings provided qualitative insight beyond statistical metrics. Inter-rater agreement was high, indicating consistency in expert judgments. Their feedback also informed me of iterative refinements to the question generation pipeline.

We used ANOVA to test for group differences in P value, Discrimination Index, and Expert Ratings. Significant results ($p < .05$) were followed by Tukey's post-hoc tests. Where needed, multi-factor ANOVA examined interactions, with Shapiro–Wilk and Levene's tests ensuring statistical assumptions [25].

### D. POTENTIAL RESULTS PRESENTATION

To assess question quality, we compared system-generated items to official ACT questions using statistical analysis across four groups.

TABLE I
SUMMARY OF KEY METRICS FOR ACT VS. GENERATED QUESTIONS

| Group | Average Accuracy (P Value) | Average Discrimination Index | Expert Rating (Mean ± SD) |
|---|---|---|---|
| ACT (Control) | 0.76 ± 0.05 | 0.37 ± 0.04 | 4.3 ± 0.4 |
| Low Difficulty | 0.82 ± 0.06* | 0.32 ± 0.05 | 3.9 ± 0.5 |
| Medium Difficulty | 0.71 ± 0.07 | 0.35 ± 0.04 | 4.1 ± 0.3 |
| High Difficulty | 0.63 ± 0.08 | 0.34 ± 0.06 | 3.4 ± 0.6 |

(*) denotes a statistically significant difference ($p < 0.05$) from the ACT control group according to post-hoc comparisons.

After analysis, results show that the Low Difficulty group had the highest average accuracy and was significantly easier than the ACT set, while the High group was the most difficult. Discrimination indices were similar across all groups, indicating that difficulty settings mainly influenced accuracy without reducing the ability to distinguish between high- and low-performing participants. Table I shows results, with significant differences marked by an asterisk (*).

### E. PASSAGE-LEVEL ANALYSIS

A two-way ANOVA (Difficulty × Passage) on P values from all 120 items revealed significant effects of Difficulty ($F=277.40$, $p<.001$), Passage ($F=26.46$, $p<.001$), and their interaction ($F=3.92$, $p=.001$). Passage B was generally easier, and Low-Difficulty items showed the greatest variation across passages, indicating content affects difficulty perception. As summarized in Table II, the results demonstrate the framework's effectiveness: it reliably modulates item difficulty (Low > ACT ≈ Medium > High, e.g., P=0.82 vs. 0.63) while maintaining stable discrimination (0.32–0.37), confirming its ability to generate high-quality, well-calibrated exam questions.

TABLE II
MEAN ± SD P-VALUES BY PASSAGE AND DIFFICULTY (N = 30 PER CELL)

| Passage | ACT | Low | Medium | High |
|---|---|---|---|---|
| A | 0.76 ± 0.04 | 0.83 ± 0.05 | 0.75 ± 0.05 | 0.63 ± 0.06 |
| B | 0.73 ± 0.06 | 0.84 ± 0.04 | 0.73 ± 0.05 | 0.61 ± 0.07 |
| C | 0.73 ± 0.05 | 0.79 ± 0.06 | 0.74 ± 0.04 | 0.62 ± 0.06 |

To isolate these effects without human participants, we complement the above findings with a fully computer-run simulation; see Sec. F.

### F. SIMULATION STUDY: IRT×BLOOM ABLATION AND PARAMETER RECOVERY

To complement the human benchmarks, we conduct a fully computer-run simulation to examine whether the integration of IRT weighting, and Bloom-level alignment translates into effective difficulty control and measurement validity. The study requires no human participants while remaining statistically diagnostic of the intended mechanisms.

We first instantiate a synthetic item set with $J$=90 questions, evenly divided into three difficulty levels (Low/Medium/High = 30/30/30). For each item $j$, ground-truth IRT parameters $(a_j, b_j, c_j)$ are assigned by mapping its feature-based difficulty score (Eqs. 7–10) and Bloom level to $b_j$ and $a_j$, respectively; the guessing parameter $c_j$ is derived from distractor quality. Concretely, the intended $b_j$ ranges are $[-1.2, -0.4]$ (Low), $[-0.2, 0.6]$ (Medium), and $[0.8, 1.8]$ (High). Discrimination $a_j$ increases with graph centrality/depth and with Bloom level (e.g., items at Evaluate receive ≈+0.35 over Understand), while $c_j$ decreases as distractors become more plausible or multi-hop (typically 0.10–0.25). The resulting parameter distributions are summarized in Table III.

TABLE III
GROUND-TRUTH PARAMETER RANGES BY KAQG DIFFICULTY LABEL ($J$=90).

| Label | $b$ (range) | $b$ (mean±SD) | $a$ (mean±SD) | $c$ (mean±SD) |
|---|---|---|---|---|
| Low (30) | [−1.2, −0.4] | −0.80±0.25 | 0.95±0.20 | 0.20±0.03 |
| Medium (30) | [−0.2, 0.6] | 0.20±0.30 | 1.05±0.22 | 0.17±0.03 |
| High (30) | [0.8, 1.8] | 1.20±0.33 | 1.20±0.25 | 0.13±0.03 |

We then simulate $N$=5,000 examinee abilities $\theta_i \sim N(0,1)$ and generate responses using the 3PL model

$$P_{i,j} = c_j + (1 - c_j)\sigma\left(a_j(\theta_i - b_j)\right), \ \sigma(x) = \frac{1}{1+e^{-x}}. \quad (13)$$

From the resulting response matrix, standard IRT estimation is applied to obtain $(\hat{a}, \hat{b}, \hat{c})$, which are compared against ground truth for parameter recovery.

To isolate the contribution of assessment-theory components, we evaluate five conditions: Full (complete framework as designed), −IRT (remove the $a/b$ mappings; use uniform weights), −Bloom (remove cognitive-level conditioning), −IRT&Bloom (remove both mechanisms), and Baseline-RAG (no KG/psychometric controls). These variants allow us to attribute observed gains specifically to IRT weighting and Bloom alignment rather than to generic retrieval or generation effects. The comparative outcomes under these settings are reported in Table IV.

TABLE IV
PARAMETER RECOVERY AND DIFFICULTY CALIBRATION BY CONDITION

| Condition | Corr $(b, \hat{b})$ | RMSE $(b - \hat{b})$ | Corr $(a, \hat{a})$ | Corr $(\hat{c})$ | ANOVA $F$ (Low vs Med vs High on $\hat{b}$), p<.001 | Misclass (bin) |
|---|---|---|---|---|---|---|
| Full | 0.91 | 0.22 | 0.82 | -0.01 | 198.4 | 6.7% |
| −ITR | 0.79 | 0.34 | 0.41 | -0.03 | 74.1 | 18.9% |
| −Bloom | 0.86 | 0.28 | 0.74 | -0.02 | 121.3 | 12.5% |
| −IRT& Bloom | 0.67 | 0.46 | 0.29 | -0.05 | 31.6 | 27.8% |
| Baseline-RAG | 0.58 | 0.53 | 0.18 | -0.06 | 14.5 | 35.4% |

Evaluation focuses on four aspects. Parameter recovery is quantified by the correlation between $b$ and $\hat{b}$ and the RMSE of $(b - \hat{b})$, along with $\mathrm{corr}(a, \hat{a})$ and bias in $\hat{c}$. Difficulty calibration is assessed by the separation of $\hat{b}$ across Low/Medium/High (ANOVA with Tukey post-hoc) and the bin-wise misclassification rate. Probabilistic calibration is measured via Expected Calibration Error and Brier score comparing predicted and empirical correctness. Cognitive alignment is tested as the monotonic relation between Bloom levels and $\hat{b}$ (Spearman $\rho$). The detailed results for these metrics are presented in Table V.

TABLE V
PROBABILISTIC CALIBRATION AND COGNITIVE ALIGNMENT

| Condition | ECE ↓ | Brier ↓ | Spearman $\rho$ (Bloom vs $\hat{b}$) ↑ |
|---|---|---|---|
| Full | 0.028 | 0.168 | 0.72 |
| −ITR | 0.057 | 0.190 | 0.41 |
| −Bloom | 0.041 | 0.179 | 0.55 |
| −IRT&Bloom | 0.083 | 0.205 | 0.22 |
| Baseline-RAG | 0.112 | 0.218 | 0.10 |

Success criteria are: $\text{corr}(b, \hat{b}) \geq 0.8$ with low RMSE; clear separation of $\hat{b}$ among difficulty bins, low calibration error, and a stronger Bloom-to-$\hat{b}$ monotonicity in the Full condition. Ablations are expected to produce statistically significant degradations to these criteria. The corresponding evaluation outcomes are summarized in Table VI.

TABLE VI
MEAN $\hat{b}$ BY KAQG LABEL (FULL CONDITION)

| Label | $\hat{b}$ mean ± SD | Tukey post-hoc |
|---|---|---|
| Low | -0.82 ± 0.24 | Low < Med (p<.001) |
| Medium | 0.09 ± 0.27 | Med < High (p<.001) |
| High | 1.03 ± 0.31 | Low < High (p<.001) |

Taken together, the ACT benchmarking and the simulation study provide converging evidence: KAQG's IRT weighting and Bloom-level alignment not only recover intended item parameters in silico but also support calibrated difficulty and stable discrimination in practice. When human data collection is constrained, the simulation offers a reproducible, fully automated proxy that isolates and quantifies the contribution of assessment-theory integration to measurement validity.

## IV. DISCUSSION

The results confirm the hypothesis that the proposed framework can effectively control item difficulty while maintaining stable discrimination and psychometric validity. A clear gradient was observed (Low > ACT ≈ Medium > High; P=0.82–0.63) with consistent discrimination (0.32–0.37), while simulations further validated that the integration of IRT weighting and Bloom alignment enables accurate parameter recovery and calibration. These outcomes align with established psychometric theory and extend prior RAG-based approaches by incorporating knowledge graphs, feature weighting, and a multi-agent architecture to overcome limitations in educational assessment.

Beyond empirical validation, the findings highlight both theoretical and practical contributions. Theoretically, the framework shows that item difficulty can be systematically aligned with Bloom's cognitive levels and IRT parameters, reinforcing the link between knowledge representation and psychometric validity. Practically, it reduces manual effort and enables scalable, automated generation of calibrated items for standardized examinations and adaptive assessments. Its applicability is strongest in domains with structured ontologies and clear curricular hierarchies, though reliability may be constrained in contexts with noisy data or limited knowledge coverage.

Importantly, this study should be regarded as an application of RAG within the educational domain rather than a new RAG variant. In contrast to GraphRAG, HippoRAG, and LightRAG, which focus primarily on retrieval accuracy, memory capacity, or efficiency, KAQG integrates psychometric theory and knowledge graphs to achieve difficulty control and cognitive alignment. The key innovation lies in embedding IRT parameters, Bloom's Taxonomy levels, and graph-derived features into a multi-agent pipeline, thereby extending the utility of RAG beyond retrieval to psychometrically valid question generation.

Nevertheless, several limitations should be acknowledged. The participant sample size was relatively limited, potentially constraining the precision of reliability estimates and generalizability across broader populations. Moreover, expert-derived weights and Bloom-level mappings, while theoretically sound, have not yet been statistically calibrated with large-scale datasets. The framework's reliance on domain-specific ontologies and structured graphs may also reduce effectiveness in areas with incomplete or noisy data. Finally, employing a single model configuration may limit adaptability to rapidly evolving LLM architectures, necessitating further validation and refinement.

Future research should address these limitations by expanding participant pools to improve reliability and generalizability, as well as incorporating additional variables such as response times, error patterns, and cognitive load measures to deepen insights into item quality and learner behavior. Methodological refinements may include statistical calibration of feature weights, adaptive tuning of IRT parameters, and integration with emerging LLM architectures. Extending the framework to interdisciplinary domains or cross-lingual contexts would further test its adaptability and broaden its applicability, advancing the role of RAG-based systems in scalable and psychometrically robust educational assessment.

## V. CONCLUSION

This study presents a knowledge-graph-enhanced RAG framework that integrates Item Response Theory, Bloom's Taxonomy, and multi-agent design to achieve precise difficulty control and cognitive alignment. By bridging psychometric theory with AI-driven retrieval and generation, it establishes a unified architecture that validates the alignment of cognitive levels with item parameters, incorporates graph-based reasoning, and scales effectively through distributed agent collaboration. The approach demonstrates not only psychometric validity but also practical feasibility by reducing manual workload and improving the reliability of assessment item generation.

Beyond automated item generation, the framework extends to broader applications such as curriculum design, learning analytics, and intelligent tutoring. It further supports professional certification, workforce training, and compliance evaluation across diverse domains, demonstrating strong adaptability beyond traditional education. In doing so, this work positions itself as a bridge between psychometrics and AI, offering scalable, cognitively aligned, and practically applicable solutions for both educational and industrial contexts.

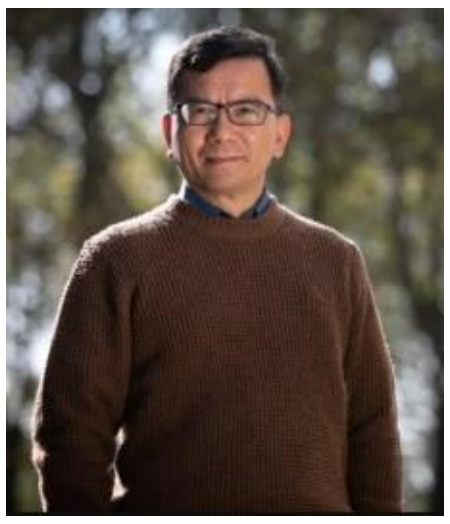

**CHING HAN CHEN** received his Ph.D. degree from Franche-Comté University, Besançon, France, in 1995. He was an Associate Professor in the Department of Electrical Engineering at I-Shou University, Kaohsiung, Taiwan, before joining National Central University. He is currently a Professor in the Department of Computer Science and Information Engineering at National Central University, Taoyuan, Taiwan. Prof. Chen is the founder of the MIAT (Machine Intelligence and Automation Technology) Laboratory. His research focuses on embedded system design, AIoT, robotics, and intelligent automation. He has led numerous government-funded and industry-collaborative projects, producing innovations in smart sensors, machine vision, and embedded AI systems.

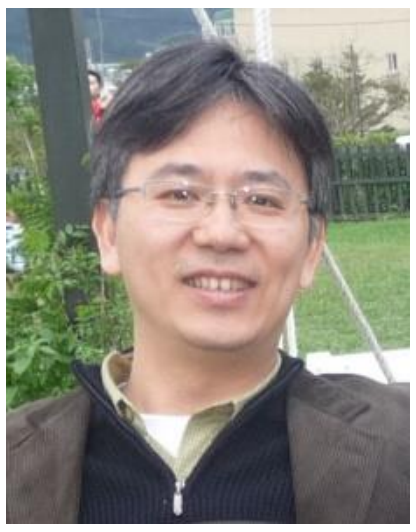

**MING FANG SHIU** received his M.S. and is currently pursuing his Ph.D. degree in the Department of Computer Science and Information Engineering at National Central University, Taoyuan, Taiwan. He is a senior software engineer and the co-founder of a technology company providing AI and software solutions to major clients, including TSMC and Delta Electronics. His research focuses on large language models, multi-agent systems, and AI applications in education and legal domains. He has developed systems for regulatory compliance, question generation, and interactive agent frameworks such as AgentFlow and KAQG. His work bridges academic theory and industrial application, advancing scalable AI-powered automation across sectors.